# Fluorescence from metallic silver and iron nanoparticles prepared by exploding wire technique

**Abdullah Alqudami\* and S Annapoorni**

Department of Physics and Astrophysics, University of Delhi, Delhi, India 110 007
 \* Author for correspondence: aalqudami@physics.du.ac.in

**Abstract**

The observation of intense visible fluorescence from silver and iron nanoparticles in different solution phases and surface capping is reported here. Metallic silver and iron nanoparticles were obtained by exploding pure silver and iron wires in pure water. Bovine serum albumin protein adsorption on the silver nanoparticles showed an enhanced fluorescence. The presence of poly-vinyl pyrrolidone polymer in the exploding medium resulted in a stabilized growth of iron nanoparticles with enhanced fluorescence intensity. The fluorescence was found to be surface /interface dependant. The fluorescence is attributed to electronic transitions among characteristic interface energy bands. The magnetic nature of iron nanoparticles was confirmed from the hysteresis measurements.

1. Introduction

Metal nanoparticles display novel physical and chemical properties due to surface effect, where most of the particle atoms are just surface atoms [1]. These novel properties have put metal nanoparticles to play an interesting role in materials technology, biomedicines, catalysis, etc. The Optical properties of metal nanoparticles are highly influenced by the preparation methods and conditions, which result in particles of various sizes; shape and surface stabilization [2, 3]. Moreover, the surface/interface interactions have their signatures in the properties investigated [4, 5]. Among all the known chemical and physical preparation methods, the exploding wire technique is one of the newest and simplest methods for producing metal nanoparticles [6, 7]. The explosion is achieved when a very high current density is applied to a thin metal wire, causing the wire to explode to very small fragments. This process involves wire heating and melting followed by wire evaporation, formation of a high-density core surrounded by low-density ionized corona, coronal compression, and fast expansion of the explosion products [8]. The exploding wire experiments were mostly employed for the generation and investigation of plasma and shock waves. In previous reports, we studied the optical properties of silver nanoparticles prepared by exploding wire technique. These silver nanoparticles showed surface plasmon absorption and plasmonic excited fluorescence [9, 10]. The fluorescence of metal clusters and thin films is well established based on Mooradian observation of photoluminescence from bulk copper and gold [11], and on photoinduced luminescence from the noble metals observed by Boyd [12]. For example, the photoinduced fluorescence from $Ag_3$ clusters during the agglomeration of silver atoms was observed to center around 500 nm as reported by Ievlev et al. [13]. Dendrimer-encapsulated silver nanodots were also observed to be fluorescent both in aqueous solutions and in films [14]. Most of the observed fluorescence was attributed to radiative recombination of an electron-hole pair between d-band and sp-conduction band above the Fermi level. Due to plasmon resonance excitation, the local field created around the nanoparticles is found to modify the observed fluorescence [15].

Recently [16], thermal growth of silver nanoparticles on soda glass was correlated with drastic changes in their photoluminescence intensity. Another model has been proposed to explain the fluorescence from silver colloidal particles as their surface was chemically changed by deposition of silver ions [17].

Silver nanoparticles in water, were found to have fluorescence peak at 465 nm. The interface electrons were proposed to be the source of the observed fluorescence [18]. Far from the fluorescence from noble metals, Fei et al. [19] observed for the first time the fluorescence from $\alpha$-$Fe_2O_3$ nanoparticles – coated with a layer of organic molecules. They attributed the fluorescence to bound - exciton emission, as also observed for other typical semiconductor nanoparticles. On the other hand, mesoscopic systems that consist of a spherical metallic core surrounded by an insulating layer were investigated theoretically, where electromagnetic decay of externally located electrons to the lower-lying unoccupied electronic states located within the metal exhibit a resonance behavior. It was shown that the probability of lowest external electrons to be found in the inner states located within the metal is increasing with the metallic core radius [20].

In the present study, we report the observation of strong visible fluorescence from silver nanoparticles in different media under plasmonic excitations, and for the first time, we have observed an intense fluorescence from iron nanoparticles under an ultraviolet excitation. Both the systems were prepared using the exploding wire technique in pure water and under the same conditions. In order to study the surface effects on the observed fluorescence, Bovine serum albumin (BSA) proteins were introduced to prepare BSA-adsorbed silver nanoparticles, while iron nanoparticles were coated with poly-vinyl pyrrolidone (PVP) polymer.

2. **Experimental methods**

Silver and iron wires (99.998 %) with diameters of 0.2 mm were exploded in double distilled water using silver and iron plates respectively, to produce silver and iron nanoparticles. 25 µM poly-vinyl pyrrolidone (PVP) polymer ($M_n$ = 40,000, Aldrich) was dissolved in double distilled water and this was used as the explosion medium for iron wires, to produce PVP- coated iron nanoparticles. These wires have been exploded by bringing the wire into sudden contact with the plate (purity 99.998 %) when subjected to a potential difference of 36 V DC. High current density was allowed to pass through these thin wires, where tensile fracture forces proportional to the square of the current cause the wires to rupture before becoming unduly softened by the Ohmic heating. The total mass being exploded was about 0.3 g in each experiment. The explosions in all the experiments were carried out under the same conditions.

BSA proteins (SRL; India) were added to the water-based silver nanoparticles solution with total concentration of 10.0 μg / ml under overnight vigorous stirring at 4 $^o$C.

Part of the silver and iron nanoparticles solutions were centrifuged at 12,000 rpm and the separated particles were washed and re-dispersed in cyclohexane.

Powder nanoparticles obtained after filtering were used for the x-ray diffraction studies that were performed with a Philips Analytical X-Ray Diffractometer type PW3710 Based using Cu- K$\alpha$ radiation (wavelength 1.54056 Å). Drops from the decanted solutions were allowed to dry on carbon-coated copper grids for electron microscopy imaging using JEOL JEM 2000EX transmission electron microscopy (TEM). Particle size distributions were performed using Photocor- F Dynamic light scattering, Photocor instruments. UV-2510PC spectrophotometer was used to record UV-Visible absorption spectra. The fluorescence emission and excitation spectra were recorded using Edinburgh Analytical Time Resolved Fluoremeter. The magnetic moment (M) of the iron wire and the water-based iron nanoparticles was measured using Lake Shore model 7304 Vibrating Sample Magnetometer (VSM).

3. Results
*3.1 X-ray diffractions*

Figure 1 (a) shows the x-ray diffractions from silver nanoparticles with respect to bulk silver wire used in the explosions. The peaks were assigned to the diffraction from the (111), (200), (220), (311) and (222) planes of FCC silver respectively. (111), and (200) peaks were observed to shift up by about 0.5 degree with respect to the corresponding peaks of the bulk. This shift might be due to slight decrease in the d spacing values of the corresponding planes. Figure 1(b) presents the x-ray pattern of iron nanoparticles with respect to bulk iron wire. The only peak observed at 44.5 degree is assigned to the diffraction from the (110) plane of BCC $\alpha$- Fe. The lattice constants calculated from these patterns are in agreement with the literature reports in both the cases; while particle size estimated using Scherrer's formula was about 63 nm in the case of silver and about 30 nm in the case of iron.

*3.2 Transmission electron microscopy*

Figures 2 (a and b) show the typical TEM images of silver nanoparticles and BSA adsorbed silver nanoparticles respectively. The adsorption of BSA proteins showed to prevent silver nanoparticles from coalescence and hence smaller agglomerated nanoparticles were observed as in figure 2(b). Figures 2 (c – e) show the TEM images of iron nanoparticles. The as prepared iron nanoparticles is shown in figure 2 (c), where particles of various sizes are observed. The size selection was maintained by centrifuging the solution of iron nanoparticles at 10,000 rpm and the results are shown in figure 2 (d), where larger particles (inset in fig 2(d)) are separated from the smaller ones. Figure 2 (e) shows the PVP-coated iron nanoparticles where the nanoparticles are stabilized by the polymer chains. The PVP coating most likely prevents iron nanoparticles from agglomerations.

### 3.3 Dynamic light scattering

Particle size distributions have been made using dynamic light scattering where He – Ne laser light (633 nm) is passed through diluted nanoparticles solutions. The size distributions of silver and iron nanoparticles are shown in figure 3 (a and b) respectively. X- axes show the particle size in μm with logarithmic scale. Silver nanoparticle sizes ranging from 10 nm to 120 nm with distribution around 70 to 100 nm were obtained, while iron nanoparticles ranging from 3 to 90 nm with mean size of 40 nm. The size distributions are in reasonable agreements with the sizes obtained using XRD and TEM.

### 3.4 UV – Visible spectra

Figure 4(a) shows the absorption spectra of silver nanoparticles and BSA proteins adsorbed silver nanoparticles respectively. The absorption spectra were recorded for the water - based solutions with the appropriate base line corrections. Surface plasmon resonance (SPR) peak of silver nanoparticles was observed at 410 nm. This is in a good agreement with the theoretical simulation of SPR using Mie's theory [10]. The SPR was enhanced through the adsorption of BSA. Moreover, the FWHM of the SPR has increased. Details on the effect of protein adsorption on the optical properties of silver nanoparticles are presented elsewhere [21].

The UV-Visible spectra of iron nanoparticles are shown in Figure 4(b). Iron nanoparticles in water showed two absorption peaks at wavelengths of 216 nm and 268 nm. These two structures are not clear in the case of PVP-stabilized iron nanoparticles. However, strong absorption is observed below 250 nm while the structure at 268 nm in the last case disappeared but with appearance of a small hump at about 360 nm. This can be assigned to small size clusters of PVP-stabilized iron nanoparticles, which is in agreement with reported absorption spectra [22, 23].

Iron nanoparticles in cyclohexane shows four absorptions at 207, 224, 252 and 280 nm. It seems that the mean absorption peaks at 216 and 268 nm of the iron nanoparticles in water have been split to four absorptions. The iron nanoparticles in cyclohexane have some selected favorite transitions, instead of broad absorption peaks in the case of water interface. The extinction coefficient of spherical iron nanoparticles having mean diameter of 10 nm, dispersed in water, has been calculated according to Mie's theory. The damping term as a function of the particle size was included in the calculations of the dielectric functions of iron nanoparticles [24]. The dielectric functions of bulk iron were taken from Johnson and Christy [25]. The correction being made by inserting a constant that includes the details of the scattering processes, and the particle size that give rise to the size dependence. This theoretical curve is in good agreements with those obtained experimentally and in more agreement with the PVP-stabilized iron nanoparticles wherein the average particle size is about 10 nm. The theoretical simulation obtained in the present work is also in agreement with the reported simulations [26].

### *3.5 Fluorescence spectra*

Figures 5 (a and b) show the fluorescence excitation and emission spectra for silver nanoparticles respectively. The excitation wavelength was fixed at 390 nm and the emission spectra were recorded for three different systems, which are water based silver nanoparticles, BSA adsorbed silver nanoparticles and cyclohexane based silver nanoparticles. Silver nanoparticles in water showed two fluorescence emissions at 490 nm and 450 nm. BSA protein adsorbed on the surface of silver nanoparticles show a blue shift of 490 nm emission peak by about 20 nm while the other emission peak at 450 nm remains unchanged. Both the fluorescence peaks get blue shifted by about 30 nm when the silver nanoparticles were dispersed in cyclohexane. Water as a medium showed a little tendency to quench the fluorescence of silver nanoparticles, while the fluorescence emission was slightly enhanced by the BSA adsorption and greatly enhanced by using cyclohexane as a dispersion medium.

For the first time we tried to study the fluorescence behavior of iron nanoparticles obtained using exploding wires in double distilled water varying the excitation wavelength from 200 nm to 420 nm. Intense fluorescence emissions were observed under the excitation of 268 nm. The fluorescence excitation and emission spectra of iron nanoparticles are shown in figure 6. The water based iron nanoparticles showed fluorescence emissions at about 590 nm, 450 nm, 350 nm and 300 nm. The fluorescence emission was fixed at 590 nm while recording the excitation spectra. Coating the iron nanoparticles with PVP polymers showed to enhance the fluorescence emissions while the mean features remains unchanged. Again, iron nanoparticles and PVP-coated iron nanoparticles were centrifuged and re-dispersed in cyclohexane. The fluorescence emissions were greatly enhanced with the observation of new emission peaks.

The excitation wavelength showed to slightly affect the fluorescence emission and this effect is shown in figure 7. The water based iron nanoparticles were excited at 224 nm and 268 nm based on the excitation spectrum shown in figure 6(a). The emission peak at 590 nm remains unchanged under both the excitations. The excitation of 224 nm resulted in the appearance of some new emissions.

Fluorescent iron nanoparticles produced using the exploding wire technique showed a superparamagnetic behavior [7]. Detailed study on the magnetic properties of these iron nanoparticles is presented elsewhere [27]. The saturation magnetization ($M_s$) values obtained for the water-based iron nanoparticles in the presence and absence of PVP were 25 emu/gm and 55 emu/gm, respectively. Assuming the lognormal size distributions of the water-based iron nanoparticles, the size of magnetic core is calculated according to the method described by *Chantrell et al.* [28], and the resultant magnetic core diameter was 4.2 nm for the PVP stabilized iron nanoparticles and 4.6 nm for the non stabilized particles. The reduced ($M_s$) values with respect to the $M_s$ value of the bulk iron wire used for the explosion (which was 222 emu/gm) are attributed to the increase in the anisotropy due to size reduction and nanoparticles surface coating.

### 4. Discussion

The exploding wire results in small clusters and evaporated atoms and ions, which are in fact the source of plasma generated during the explosion process. The resulted fragments will be too dense and give rise to high filling factor. Coagulation aggregates are easily formed due to Van der Waals forces or bonding between two clusters approaching each other. The presence of some amounts of stabilizers such as PVP polymer in the explosion medium can terminate the aggregation. Water as a medium will provide fast cooling for the generated particles with energetically favored shapes. Moreover, the high surface area per unit mass of the metal nanoparticle enhances the surface activity and tends to react with hydroxide molecules. Metal nanoparticle surface, in the absence of polymers, thus may get oxidized. This process results in forming some sort of oxide layer. The coating process occurs during and after the particle growth and agglomerations. The oxidation of iron nanoparticles is reflected through the increasing of measured pH value from 6.4 for the distilled water used to 7.14 for the water-based iron nanoparticles solution. PVP coated iron nanoparticles solution has a pH value of 6.82. Thus PVP avoids the oxidation of the nanoparticles.

Metallic nanoparticles in solution contain a few square meters of interface and hence their optical properties are very strongly influenced (with respect to both its position and shape) by chemical modifications to the surface.

When metals are present in finely dispersed form, the density of electronic states will be lower, while the density of electrons at the surface will be higher due to the short diffusion time on a femtosecond scale. The thermalized time is on a picosecond scale, thus it is the ultra-fast dynamic process, which is the reason behind the high density of electrons at the surface of nanoscale metallic particles. The molecular adsorption to the particle surface is also associated with electron density donation to the particle surface [29]. However, Oxygen molecules of the water have strong affinity to bind with electrons. This effect can be clearly observed by looking at the surface plasmon resonance of the silver nanoparticles and BSA-adsorbed silver nanoparticles. The SPR signal as well as the fluorescence has been enhanced after the adsorption process. The SPR is the property of the surface free electrons; more free electrons give more SPR signal. BSA proteins adsorbed to the silver nanoparticles preventing the electrons from binding with oxygen molecules. Moreover, the agglomerated silver particles were break to small particles during the stirring and adsorption process. This process results in much more numbers of small nanoparticles, each of them will have surface free electrons contributed to the total fluorescence signal. Study on palladium particles stabilized by Poly vinyl pyrrolidone (PVP) showed that the electron density at the particle surface is enhanced by the polymer macroligand [30]. However, the electrons donated or prevented by a chemisorbed molecule are chemically bound to the metal surface atoms and are probably squeezing the free electrons at the surface into a smaller region, which leads to an increased electron density. Interface electrons then may form characteristic interface electron energy band (IEEB). The IEEB is the characteristic of the core material while the material / medium interface play an important role in the selection process. Electromagnetic excitation makes electrons at the higher occupied energy level (HOEL) undergo electronic transitions to the lower unoccupied energy level (LUEL) or excited states. Figure 8 shows the proposed diagram of the transitions occurring in the IEEB. Most of the electrons in the excited state III may undergo non-radiative decay (may be electron phonon interaction) to excited state II. Some of the electrons in the excited state II undergo fluorescence transitions (electron hole recombination) to the highest occupied energy level in the IEEB. Still some of the electrons in the excited state II may decay non-radiatively to lower excited states and then transition through energy decay producing fluorescence emission at lower energies (high wavelengths).

Iron nanoparticles stabilized by PVP polymer showed an enhanced fluorescence emission as expected since the electron density at the particle surface is enhanced. Unlike water, cyclohexane as a medium will not affect the surface free electrons at all and hence the fluorescence intensity is highly enhanced from both silver and iron nanoparticles. However, the selection rule of the electronic transitions is changing. This can be observed through the shift of the fluorescence peaks depending on the medium (environment).

Similar explanation was also reported in systems such as silver nanoparticles in water phase [18], and coated $\alpha$ – $Fe_2O_3$ nanoparticles [19].

Regardless of the electrons at the interface, the emissive center can be the metallic nanoparticle core as it was concluded that external electrons can decay electromagnetically to internal electronic states located within the metallic core [20]. This process of course, depends on the metallic core size. The emissive centers can be something else as well. In the case of silver and iron nanoparticles prepared by the exploding wire technique, the emission of plasma light during the explosion process is a clear indication to the generation of ions. Some of these ions may get adsorbed to the surface of the formed nanoparticles and they became the emissive centers. Similar proposal was put to explain fluorescence emission from silver nanoparticles in solution exposed to silver ions [17]. The interaction between the emissive centers and the interface environment also has its signature on the observed fluorescence.

Indeed, the explosions were carried out in water and the fluorescence was recorded in air. These may give rise to partial surface oxidation. One suggests that partial oxidations of metal nanoparticles lead to the formations of metal oxide clusters on the metal nanoparticle surface. These clusters at the surface are photo-activated by light excitations and might give rise to the observed fluorescence. Photo-activated fluorescence has been previously observed from small silver clusters [31].

## 5. Conclusion

Although the observation of fluorescence from noble metals nanoparticles and nanoclusters is not new, models put to explain the behavior are still not clear enough and, most of the time, are not identical. Further, studying the fluorescence of metallic nanoparticles should be extended to other applicable systems. Here the fluorescence of iron nanoparticles with their magnetic nature is observed for the first time. We report the fluorescence from silver nanoparticles prepared by the same technique using the same mediums. The fluorescence of silver nanoparticles more or less agrees with literatures. In fact the fluorescence from iron nanoparticles under investigation is much more than that from silver nanoparticles.

The possible explanation of these emissions is the creation of surface/interface energy bands where the probability of some allowed transitions is increased. Fluorescent magnetic nanoparticles are totally new discovery. The metal nanoparticles obtained may find applications in the field of water purifications and biomedicines.

**Acknowledgments**

We would like to acknowledge Dr. N C Mehra and Mr. Raman of University Science Instrumentation Center (USIC), and Mr. P C Padmakshan of the Geology Department, University of Delhi for recording TEM, UV-Visible, and XRD respectively, Dr. N K Chaudhary of Institute for Nuclear Medicines and Allied Sciences (INMAS), Delhi for recording the fluorescence spectra. Dr. R K Kotnala of National Physical Laboratory, Delhi for carrying out the magnetization measurements. We would also like to acknowledge the Department of Science and Technology (DST), India for the funding through the project (SR/S5/NM-52/2002) from the Nanoscience and Technology Initiative programme.

**Figure captions:**

Figure 1. X-ray diffraction of nanoparticles obtained by exploding wire. (a) silver nanoparticles (b) iron nanoparticles. The diffraction for the bulk silver and iron wires used are also shown for comparison.

Figure 2. Transmission Electron Micrographs of (a) filtered silver nanoparticles, (b) BSA - silver nanoparticles, (c) as obtained iron nanoparticles, (d) remaining iron nanoparticles after centrifuging at 10,000 rpm (inset showed the collected particles), and (e) PVP – coated iron nanoparticles.

Figure 3. Particle size distribution of as obtained (a) silver nanoparticles, (b) iron nanoparticles, estimated by Dynamic light scattering.

Figure 4. UV-Visible spectra of (a) silver nanoparticles and (b) iron nanoparticles.

Figure 5. Fluorescence spectra of silver nanoparticles (a) excitation spectra of silver nanoparticles in water at emission wavelength of 490 nm and (b) emission spectra of silver nanoparticles in water, BSA adsorbed silver nanoparticles in water, and silver nanoparticles in cyclohexane, at excitation wavelength of 390 nm.

Figure 6. Fluorescence spectra of iron nanoparticles (a) excitation spectra of iron nanoparticles in water at emission wavelength of 590 nm and (b) emission spectra of iron and PVP coated iron nanoparticles both in water and in cyclohexane mediums, at excitation wavelength of 268 nm.

Figure 7. Fluorescence emission of iron nanoparticles in water at two different excitations.

Figure 8. The propsed diagram of the transitions occurring in the interface electron energy band (IEEB).

**List of figures**

Figure 1

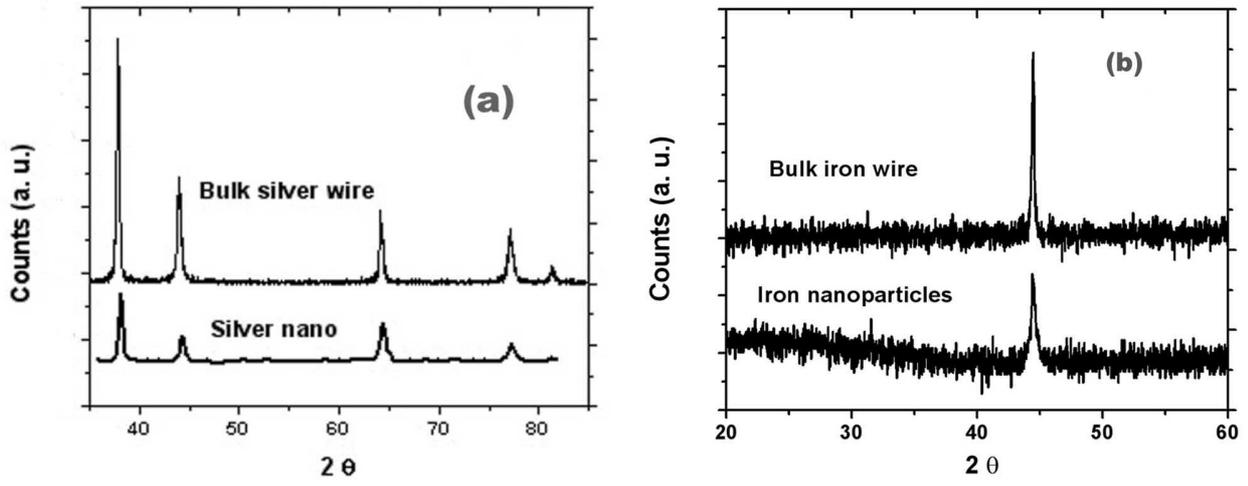

...........................................................................................................................

Figure 2

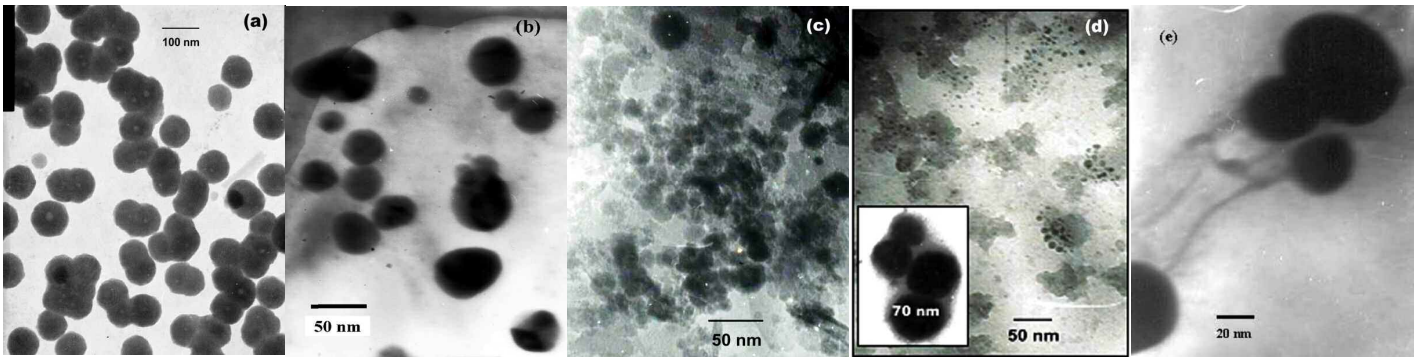

Figure 3

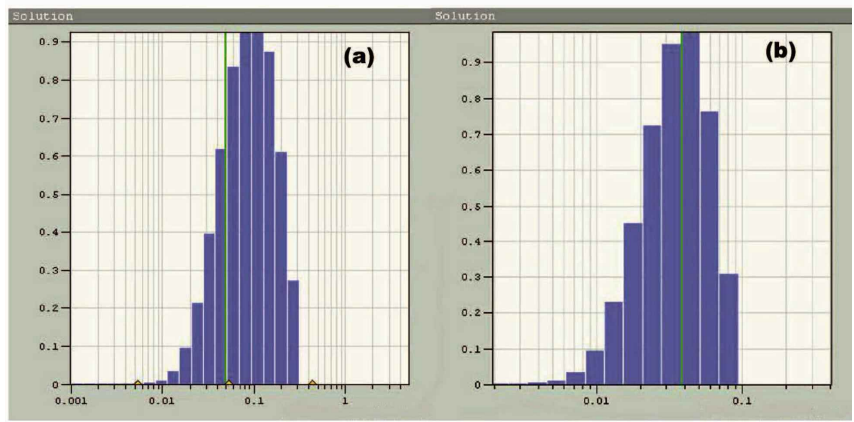

..................................................................................................................

Figure 4

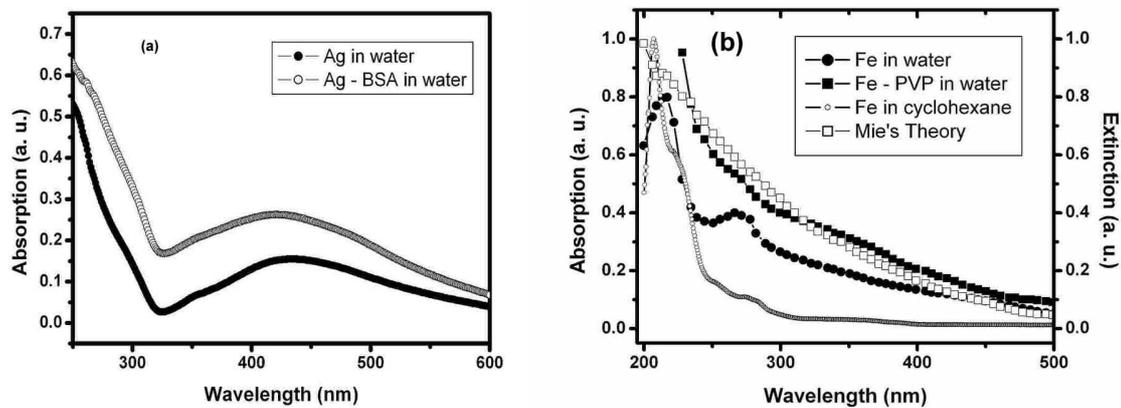

Figure 5

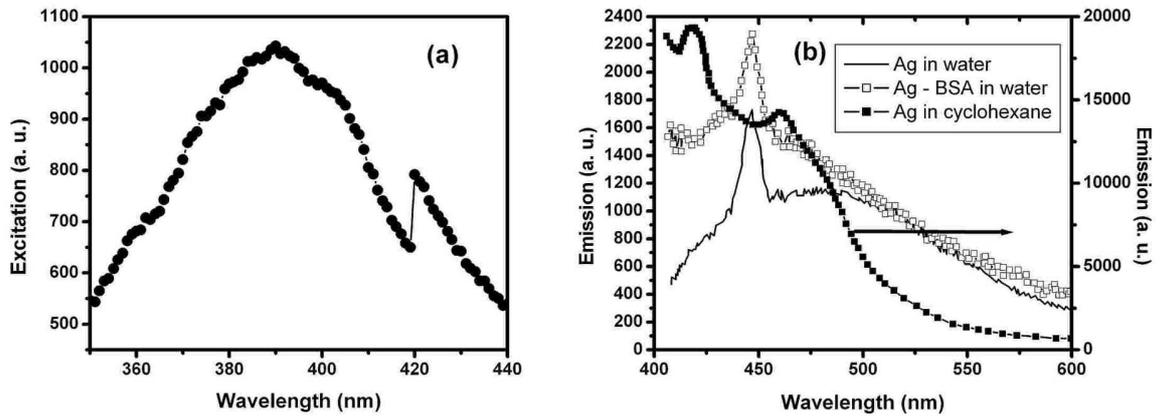

Figure 6

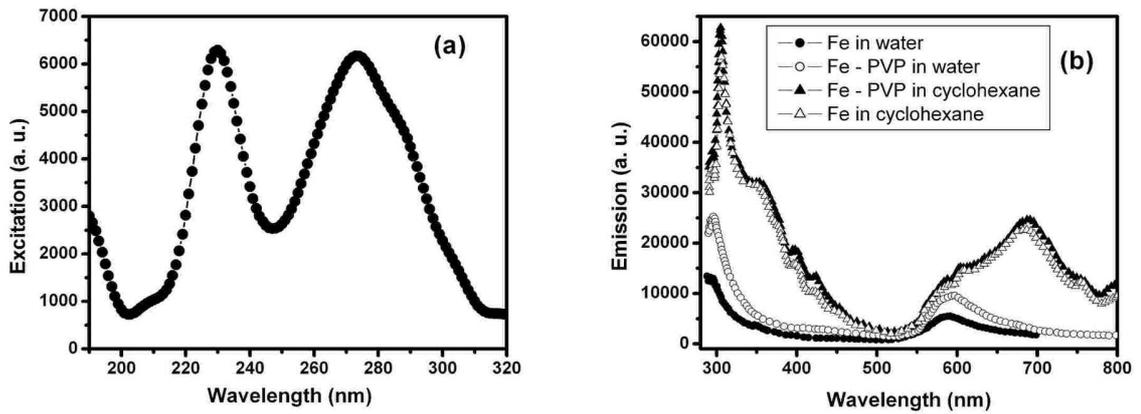

Figure 7

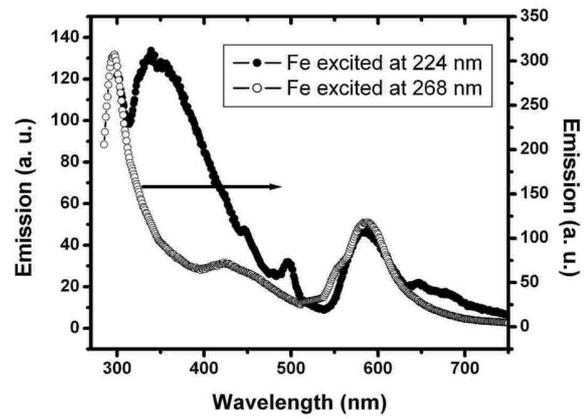

..................................................................................................................

Figure 8

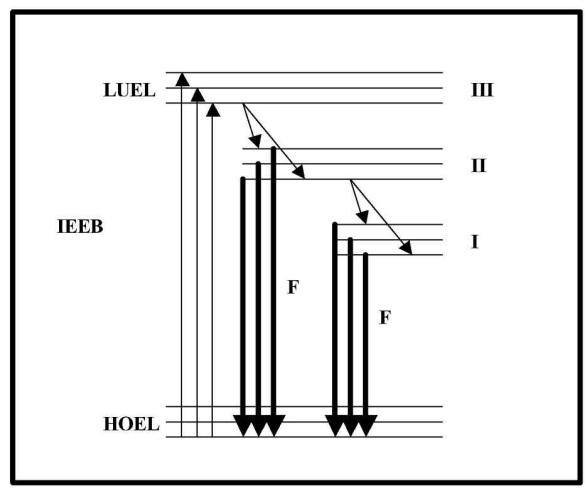